\documentstyle[12pt,epsfig]{article}
\begin{document}
\baselineskip 20pt        
\noindent
\hspace*{13cm}
hep-ph/9710428\\
\noindent
\hspace*{13cm}
SNUTP 97-124\\  
\noindent
\hspace*{13cm}
YUMS 97-027\\ 

\vspace{1.8cm}

\begin{center}
{\Large \bf
Measuring $ \left |  \frac{V_{td}}{V_{ub}} \right |$ \\
through $B \rightarrow M \nu \bar\nu~ (M=\pi,K,\rho,K^*)$ 
decays }\\

\vspace*{1cm}

{\bf T. M. Aliev $^{a,}$\footnote{taliev@rorqual.cc.metu.edu.tr}} 
and 
{\bf C. S. Kim $^{b,}$\footnote{kim@cskim.yonsei.ac.kr,~ 
cskim@kekvax.kek.jp,~ http://phya.yonsei.ac.kr/\~{}cskim/}}\\

\vspace*{0.5cm}

 $a:$ Physics Department, Girne American University, \\
        Girne, Cyprus \\
\vspace{5mm}

 $b:$ Physics Department, Yonsei University, Seoul 120-749, Korea \\
       
\vspace{1.0cm}
 
(\today)
\vspace{0.5cm}

\end{center}        

\begin{abstract}
\vspace{0.2cm}

\noindent
We propose a new method for precise determination of $ \left |
\frac{V_{td}}{V_{ub}} \right | $ from the ratios of branching ratios $\frac{
{\cal B}(B \rightarrow \rho \nu \bar \nu )} { {\cal B}(B \rightarrow
\rho l \nu )}$ and $\frac{ {\cal B}(B \rightarrow \pi \nu \bar \nu )}
{ {\cal B}(B \rightarrow \pi l \nu )}$. These ratios depend only on
the ratio of the Cabibbo-Kobayashi-Maskawa (CKM) elements $ \left |
\frac{V_{td}}{V_{ub}} \right | $ with little theoretical
uncertainty, when very small isospin breaking effects are neglected.
As is well known, $ \left | \frac{V_{td}}{V_{ub}} \right | $ equals to
$ \left( \frac{\sin \gamma }{\sin \beta } \right)$ for the CKM version of
CP-violation within the Standard Model. We also give in detail
analytical and numerical results on the differential decay width
$\frac {d\Gamma (B\rightarrow K^* \nu \bar \nu )}{dq^2}$ and the ratio
of the differential rates $\frac{d{\cal B} (B \rightarrow \rho \nu
\bar \nu )/dq^2} {d{\cal B}(B \rightarrow K^* \nu \bar \nu )/dq^2}$ as
well as $\frac{ {\cal B}(B \rightarrow \rho \nu \bar \nu )} { {\cal
B}(B \rightarrow K^* \nu \bar \nu)}$ and $\frac{ {\cal B}(B
\rightarrow \pi \nu \bar \nu )} { {\cal B}(B \rightarrow K \nu \bar \nu)}$.

\end{abstract} 
 
\newpage

\section{Introduction}

The determination of the elements of the Cabibbo-Kobayashi-Maskawa
(CKM) matrix is one of the most important issues of quark flavor
physics. The precise determination of $V_{td}$ and $V_{ub}$ elements
has principal meaning, since they are solely responsible for the
origin of CP violation in the CKM version of CP-violation within the
Standard Model (SM).  Furthermore, the accurate knowledge of these
matrix elements can be useful in relating them to the fermion masses
and also in searches for hints of new physics beyond the SM.
Therefore, strategies for the accurate determination of $V_{td}$ and
$V_{ub}$ are urgently required.  In the existing literature, we can find
proposals of different methods for precise determination of $V_{ub}$
and $V_{td}$ from inclusive and exclusive, semileptonic and non
leptonic decays of $B$ meson (see \cite{ref1} for a recent review).

The quantity $ \left| V_{ub} / V_{cb} \right| $ has been historically
measured by looking at the endpoint of the inclusive lepton spectrum
in semileptonic $B$ decays, or from the exclusive semileptonic decays
$B \rightarrow \rho l \nu $.  It has been suggested that the
measurements of hadronic invariant mass spectrum
\cite{kim-ko,bigi-wise} as well as hadronic energy spectrum
\cite{bouzas} in the inclusive $B \rightarrow X_{c(u)} l \nu$ decays
can be useful in extracting $|V_{ub}|$ with better theoretical
understanding. The measurement of the ratio $|V_{ub}/V_{ts}|$ from the
differential decay widths of the processes $B \rightarrow \rho l \nu$
and $B \rightarrow K^* l \bar l$ by using $SU(3)$-flavor symmetry and
heavy quark symmetry has also been proposed \cite{sanda}.  There
has also been recent theoretical progress on the exclusive 
$b \rightarrow u$ semileptonic decay form factors using HQET-based
scaling laws to extrapolate the form factors from semileptonic $D$
meson decays \cite{hqet-based}.  The element $V_{td}$ can be extracted
indirectly from $B_d - \overline{B_d}$ mixing. 
However, in $B_d - \overline{B_d}$ 
mixing the large uncertainty of hadronic matrix
elements prevents one from extracting $V_{td}$ with good accuracy. A better
extraction of $ \left |V_{td} / V_{ts}\right |$ can be made if
$B_s - \overline{B_s}$ mixing is measured as well, since the ratio
$(f^2_{B_d}B_{B_d})/(f^2_{B_s}B_{B_s})$ can be determined much better.
Another method to determine $ \left | V_{td} / V_{ts} \right |$ comes
from the analysis of the invariant dilepton mass distributions of 
$B \rightarrow X_{d,s} l^+ l^-$ decays \cite{ref13}. 
An interesting strategy
for measuring $|V_{td} / V_{us}|$ was proposed in \cite{ref14}, 
which uses isospin symmetry to relate the decay 
$K^+ \rightarrow \pi^+ \nu \bar \nu $ 
to the well measured decay $K^+ \rightarrow \pi^0 l \nu $.
 
In this work we propose a new method to determine the ratio $ \left |
V_{td} /V_{ub} \right | $ from an analysis of exclusive $B
{\rightarrow} M \nu \overline{\nu}$ decays, where $M$ means
pseudoscalar $\pi, K$ and vector $\rho, K^*$ mesons.
The inclusive $B \rightarrow X_q \nu \overline{\nu}$ decay is
theoretically very clean because of the absence of any long distance
effects and very small QCD corrections $(\sim 3\%)$ \cite{ref1,ref15},
and is therefore practically free from the scale ($\mu$) dependence.
However, in spite of such theoretical advantages, it would be
very difficult to detect this inclusive decay in experiments
because the final state contains two missing neutrinos and (many) hadrons.
 
This paper is organized as follows. In Section 2 we give the 
necessary theoretical framework to describe 
$B \rightarrow M \nu \bar \nu$  decays. 
In Section 3 we study the ratios of branching fractions
\begin{displaymath}
 {\cal B}(B \rightarrow \rho \nu \bar \nu ) /
 {\cal B}(B \rightarrow \rho l \nu )~~~ {\rm and}~~~  
 {\cal B}(B \rightarrow \pi \nu \bar \nu ) / 
 {\cal B}(B \rightarrow \pi l \nu ) .
\end{displaymath}
We also study the $q^2$
dependence of the differential decay rate of $B\rightarrow K^* \nu\bar\nu$,
and the ratio of the differential decay rates
\begin{displaymath} 
\frac{ d \Gamma (B \rightarrow \rho \nu \bar\nu) }{dq^2} /
\frac{ d \Gamma (B \rightarrow K^* \nu \bar \nu) }{dq^2},
\end{displaymath}
as well as 
\begin{displaymath}
 {\cal B}(B \rightarrow \rho \nu \bar \nu ) /
 {\cal B}(B \rightarrow K^* \nu \bar \nu )~~~ {\rm and}~~~  
 {\cal B}(B \rightarrow \pi \nu \bar \nu ) / 
 {\cal B}(B \rightarrow K \nu \bar \nu ) .
\end{displaymath}
Section 4 is devoted to a discussion of our results and conclusion.
 
\section{\bf Theory of 
$B \rightarrow M \nu \bar\nu~(M=\pi,K,\rho,K^*)$ decays} 

In the Standard Model (SM), the process $B \rightarrow M \nu \bar \nu$ 
is described at quark level by the $b \rightarrow q \nu \bar \nu$ transition, 
and receives contributions from $Z$-penguin and box diagrams, where dominant
contributions come from intermediate top quarks. The effective Hamiltonian
responsible for $b \rightarrow q \nu \bar \nu$
decays is described by only one Wilson
coefficient, namely $C^{\nu}_{10}$, and its explicit form is
\begin{equation} 
H_{eff} =\frac{G_{_F} \alpha}{2 \pi \sqrt {2}}~ C_{10}^{\nu} 
~(V_{tb}V_{tq}^*)~
\overline{q}{\gamma}^{\mu}(1-{\gamma}_5)b~
\overline{\nu}{\gamma}_{\mu}(1-{\gamma}_5)\nu , 
\label{(1)} 
\end{equation} 
where $G_{_F}$ is the Fermi constant, $\alpha$ is the fine structure
constant (at the $Z$ mass scale), and $V_{ij}$ are elements of the CKM
matrix.  In Eq. (1), the Wilson coefficient $C^{\nu}_{10}$ has the
following form, including ${\cal O}({\alpha}_s)$ corrections:
\begin{equation} 
C_{10}^{\nu} =\frac{X(x_t)}{ \sin^2 \theta_w } ,
\label{(2)} 
\end{equation}
where  
\begin{equation} 
X(x_t)=X_0(x_t) + \frac{\alpha_ s}{4 \pi} X_1(x_t) . 
\label{(3)} 
\end{equation}
In Eq. (3),
\begin{displaymath} 
X_0(x_t) =\frac{x_t}{8} \left[\frac{x_t+2}{x_t-1} + 
  \frac{3x_t-6}{(x_t-1)^2}\ln(x_t) \right] 
\end{displaymath} 
is the Inami-Lim function \cite{ref16}, and
\begin{eqnarray}
X_1(x_t) &=& \frac{4x_t^3-5x_t^2-23x_t}{3(x_t-1)^2}-
\frac{x_t^4+x_t^3-11x_t^2+x_t}{(x_t-1)^3}\ln(x_t)  \nonumber \\
&+& \frac{x_t^4-x_t^3-4x_t^2-8x_t}{2(x_t-1)^3} \ln^2(x_t)+ 
\frac{x_t^3-4x_t}{(x_t-1)^2}Li_2(1-x_t) 
+8x_t\frac{\partial X_0(x_t)}{\partial x_t}\ln(x_{\mu}) ,  \nonumber
\end{eqnarray}
where
\begin{eqnarray}
&{}& Li_2(1-x_t) = \int_1^{x_t} dt \frac{\ln(t)}{1-t},  \nonumber
\end{eqnarray}
is the Spence function, and
\begin{eqnarray}
x_t = \frac{{m_t}^2}{m_{_W}^2} ,~~~{\rm and}~~~x_{\mu} = 
\frac{{\mu}^2}{m_{_W}^2} .
\nonumber
\end{eqnarray}
Here $\mu$ describes the scale dependence when leading QCD corrections
are taken into account.  The term $X_1(x_t)$ is calculated in
Ref. \cite{ref15}.  The presence of only one operator in the effective
Hamiltonian makes the process $ b \rightarrow q\nu\bar\nu$ very
attractive, because the estimated theoretical uncertainty is related
only to the value of the Wilson coefficient $C^{\nu}_{10}$ ($i.e.$ the
uncertainty due to the top quark mass), contrary to the 
$b \rightarrow q l^+ l^-$ decay, 
where the uncertainties are described by three
independent Wilson coefficients, $C_7$, $C_9$ and $C_{10}$.  Another
favorable property of this decay is the absence of any long distance
effects, which make the $b \rightarrow q l^+l^-$ process considerably more
complicated. In spite of such theoretical advantages, in practice
the inclusive channel $B \rightarrow X_q \nu \bar \nu$ would be very 
difficult to detect in experiments. Only exclusive channels, namely 
$B \rightarrow M \nu \bar {\nu}$, may be studied experimentally.

At this point we consider the problem of computing the matrix elements
of the effective Hamiltonian (1) between $B$ and $M$ states. This
problem is related to the non-perturbative sector of QCD, and it can
be solved only by using non-perturbative methods. The matrix element
$<M{\mid} H_{eff}{\mid}B>$ has been investigated through different
approaches, such as chiral perturbation theory \cite{ref17}, three
point QCD sum rules \cite{ref18}, relativistic quark model by the
light front formalism \cite{ref19}, effective heavy quark theory
\cite{ref20}, light-cone QCD sum rules \cite{ref21}-\cite{ref23},
$etc$.

The hadronic matrix elements for $B\rightarrow P \nu \bar {\nu}$ 
($P$ is a pseudoscalar meson, $\pi$ or $K$) 
decays can be parametrized in
terms of the form-factors $f^P_+(q^2)$ and $f^P_-(q^2)$ in the
following way;
\begin{equation}
<P(p_2){\mid}\overline{q}{\gamma}_{\mu}(1-{\gamma}_5)b{\mid}B(p_1)>=
p_{\mu}f_+^P(q^2)+q_{\mu}f_-^P(q^2) ,
\label{(4)}  
\end{equation}
where $p=p_1 + p_2$ and $q=p_1 - p_2$. 
For $B \rightarrow V \nu \overline{\nu} $ ($V$ is the vector
$\rho$ or $K^*$ meson) decays, the hadronic matrix element can be
written in terms of five form-factors:
\begin{eqnarray}
 <V(p_2, \varepsilon )|\overline{q}{\gamma}_{\mu}(1 &-& {\gamma}_5)b|
B(p_1)> = - {\varepsilon}_{\mu\nu \alpha \beta} {\varepsilon}^{*\nu}
p_2^{\alpha}q^{\beta} \frac{2V(q^2)}{m_{_B} +
m_{_V}} -i[{\varepsilon}_{\mu}^*(m_{_B}+m_{_V})A_1(q^2) \nonumber \\
&-& ({\varepsilon}^*q)(p_1+p_2)_{\mu}\frac{A_2(q^2)}{m_{_B}+m_{_V}}-q_{\mu}
({\varepsilon}^*q)\frac{2m_{_V}}{q^2}(A_3(q^2)-A_0(q^2))]
\label{(5)}
\end{eqnarray}
with condition
\begin{equation} 
A_3(q^2=0)=A_0(q^2=0) . 
\label{(6)} 
\end{equation}
Note that after using the equations of motion the form-factor $A_3(q^2)$ 
can be written as a linear combination of the form-factors $A_1$ and $A_2$
(for more details see the first reference in \cite{ref18}):
\begin{equation} 
A_3(q^2)=\frac{1}{2m_{_V}}[(m_{_B}+m_{_V})A_1(q^2)-(m_{_B}-m_{_V})A_2(q^2)].
\label{(7)}
\end{equation}

In Eq. (5), $\varepsilon_\mu,~p_2$ and $m_{_V}$ 
are the polarization 4--vector,
4--momentum and mass of the vector particle, respectively.
Using Eqs. (1), (4) and (5), 
and after performing summation over vector meson
polarization and taking into account the number of light neutrinos
$N_{\nu}=3$, we have:
\begin{equation}
\frac{d{\Gamma}}{dq^2}(B^{\pm} \rightarrow P^\pm {\nu} \bar \nu)
=\frac{G_{_F}^2{\alpha}^2}{2^8{\pi}^5}{\mid}V_{tq}V^*_{tb}{\mid}^2
{\lambda}^{3/2}(1,r_{_P},s)m_{_B}^3{\mid}C_{10}^{\nu}{\mid}^2{\mid}f_p^+
(q^2){\mid}^2
\label{(8)} 
\end{equation}
and
\begin{eqnarray}
\frac{d{\Gamma}}{dq^2}(B^{\pm} \rightarrow V^\pm {\nu} \bar \nu)
&=& \frac{G_{_F}^2{\alpha}^2}{2^{10}{\pi}^5}{\mid}V_{tq}V^*_{tb}{\mid}^2
{\lambda}^{1/2}(1,r_{_V},s)m_{_B}^3{\mid}C^\nu_{10}{\mid}^2 \\
&\times& \Bigg( 8{\lambda}s
\frac{V^2}{(1+\sqrt{r_{_V}})^2}+\frac{1}{r_{_V}}\bigg[{\lambda}^2
\frac{A_2^2}{(1+\sqrt{r_{_V}})^2} \nonumber \\
&+& (1+\sqrt{r_{_V}})^2({\lambda}+12 r_{_V} s)A_1^2 
-2{\lambda}(1- r_{_V} - s) Re(A_1A_2) \bigg] \Bigg) .. \nonumber  
\end{eqnarray}
In Eqs. (8) and (9), ${\lambda}(1,r_{_M},s)$ is the usual triangle function
\begin{displaymath}
{\lambda}(1,r_{_M},s)=1+r_{_M}^2 +s^2-2r_{_M}-2s-2r_{_M}s \quad 
~~{\rm with}~~
r_{_M}=\frac{m_{_M}^2}{m_{_B}^2}, \quad s=\frac{q^2}{m_{_B}^2} .
\end{displaymath}
Similarly, calculations for the $B^{\pm} \rightarrow M^0 e^{\pm} {\nu}$ 
decay lead to the following results:
\begin{equation}
\frac{d\Gamma}{dq^2}(B^{\pm} \rightarrow P^0 e^{\pm}{\nu}) 
=\frac{G_{_F}^2}{192 \, {\pi}^3}{\mid}V_{qb}{\mid}^2
{\lambda}^{3/2}(1,r_{_P},s)m_{_B}^3 {\mid}f_p^+(q^2){\mid}^2 ,
\end{equation}
and
\begin{eqnarray}
\frac{d\Gamma}{dq^2}(B^{\pm} \rightarrow V^0 e^{\pm}{\nu}) 
&=& \frac{G_{_F}^2{\mid}V_{qb}{\mid}^2{\lambda}^{1/2}m_{_B}^3}{768{\pi}^3}
 \Bigg( 8{\lambda}s
\frac{V^2}{(1+\sqrt{r_{_V}})^2}+\frac{1}{r_{_V}}\bigg[{\lambda}^2
\frac{A_2^2}{(1+\sqrt{r_{_V}})^2} \nonumber \\
&+& (1+\sqrt{r_{_V}})^2({\lambda}+12r_{_V}s)A_1^2-2{\lambda}
   (1-r_{_V}-s)Re(A_1A_2) \bigg] \Bigg) .
\label{(10)} 
\end{eqnarray}

\section{Numerical analysis}

In deriving Eqs.(8)--(11), we set the masses of $M^+$ and $M^0$ equal
and the electron mass is neglected. Using isospin symmetry the
branching ratio for 
$B^{\pm} \rightarrow {\rho}^{\pm}\overline{\nu}\nu $ 
can be related to that for 
$B^{\pm} \rightarrow {\rho}^{0}\overline e\nu $.  
It is clear that their ratio is independent of form-factors,
$i.e.$ free of hadronic long--distance uncertainties in the limit
$m_{\rho^\pm}=m_{\rho^0}$.  Corrections to the strict isospin
symmetry, which come from phase space factors due to the difference of
masses of ${\rho}^{\pm}$ and ${\rho}^{0}$, isospin violation in the
$B\rightarrow {\rho}$ form-factors and electromagnetic radiative
corrections to the $b\rightarrow q e {\nu}$ transition,
are all small. In the following discussions we
shall neglect these small  isospin violation effects.
Also note that these corrections for $K \rightarrow \pi$ transition have 
been calculated in \cite{ref24} and found to be small, $\sim 5\%.$

Now we relate the branching ratio
${\cal B}(B^{\pm} \rightarrow {\rho}^{\pm}\bar{\nu}\nu)$ with 
${\cal B}(B^{\pm} \rightarrow {\rho}^0e^{\pm}\nu)$.
From Eqs. (9) and (11), we have
\begin{equation} 
\frac{{\cal B}(B^{\pm} \rightarrow {\rho}^{\pm}\overline{\nu}\nu)}
{{\cal B}(B^{\pm} \rightarrow
{\rho}^0e^{\pm}\nu)}=6\frac{{\alpha}^2}{4{\pi}^2}{\mid}C_{10}^{\nu}{\mid}^2
\left| \frac{V_{td}}{V_{ub}}\right|^2 .
\label{(11)} 
\end{equation}
Here the numerical factor 6 comes from the  number of light neutrinos,
and isospin symmetry relation between the form-factors of
$B^{\pm}\rightarrow {\rho}^{\pm} $ and  $B^{\pm}\rightarrow {\rho}^0$. 
In Eq. (12), we also put ${\mid}V_{tb} {\mid}=1$.
{}From Eq. (12), we  get
\begin{equation} 
\left | \frac{V_{td}}{V_{ub}} \right |^2 
= \frac{1}{6C} \frac{{\cal B}_{exp}(B^{\pm}\rightarrow 
{\rho}^{\pm}{\nu}\bar\nu)}
{{\cal B}_{exp}(B^{\pm}\rightarrow {\rho}^0e^{\pm}\nu)}
=\left( \frac{\sin \gamma }{\sin \beta} \right)^2  ,
\label{(12)}
\end{equation}
where 
\begin{displaymath}
C=\frac{{\alpha}^2}{4{\pi}^2} \left |C_{10}^{\nu} \right |^2 .
\end{displaymath}
The second relation in (13) holds only for the CKM version of 
CP-violation within the SM.
 
{}From Eq. (13), we can see that measurements of the ratio of the
branching fractions allow to determine the ratio of $\sin\gamma$ and
$\sin\beta$. Up to now\footnote{
See also the recent work \cite{kly} on the simultaneous determination of 
$\sin\alpha$ and $\sin\gamma$ from 
$B^0_{d,s} \rightarrow K, \pi$ decays.}, 
various methods for measuring each angle
separately have been proposed, $e.g..$, the angle $\beta$ will be
measured from $B \rightarrow J/\psi K_s$ decay with high accuracy, and
angle $\gamma$ is from the charged $B$ decay 
$B^\pm \rightarrow D K^\pm$ with larger uncertainty.  
As follows from Eq. (13), one can
measure the angle $\gamma$ with small theoretical uncertainty, 
if $\sin \beta $ is measured independently with high accuracy.
The following relations will also be useful 
for extracting the phase angle $\gamma$ precisely:
\begin{equation} 
\frac{{\cal B}(B^0 \rightarrow {\rho}^0{\nu} \bar\nu )}
{{\cal B}(B^0 \rightarrow {\rho}^{\pm} e^{\mp} \nu )}
=\frac{3}{2} \left ( \frac{\sin {\gamma}}{\sin {\beta}} \right )^2  C ,
\label{(14)} 
\end{equation}
\begin{equation} 
\frac{{\cal B}(B^{\pm} \rightarrow {\pi}^{\pm}{\nu}\bar\nu)}
{{\cal B}(B^{\pm}\rightarrow {\pi}^0 e^{\pm} \bar\nu )}
=6 \left ( \frac{\sin {\gamma}}{\sin {\beta}} \right )^2 C  ,
\label{(13)} 
\end{equation}
\begin{equation} 
\frac{{\cal B}(B^0 \rightarrow {\pi}^0 {\nu} \bar\nu)}
{{\cal B}(B^0 \rightarrow {\pi}^\pm e^{\mp} \nu )}
= \frac{3}{2} \left ( \frac{\sin {\gamma}}{\sin {\beta}} \right )^2 C ,
\label{(13-1)} 
\end{equation}
and
\begin{equation} 
\frac{{\cal B}(B^\pm \rightarrow K^{*\pm}{\nu} \bar\nu )}
{{\cal B}(B^\pm \rightarrow {\rho}^0 e^{\pm} \nu )}
\approx 6 \left| \frac{V_{ts}}{V_{ub}} \right|^2 C .
\label{(141)} 
\end{equation}
In derivation (14)-(17), we assumed that the mass of charged and neutral
final states mesons are equal. 
  
%
\begin{figure}[tb]
\vspace*{-5cm}
\hspace*{-2cm}
\centerline{\epsfig{figure=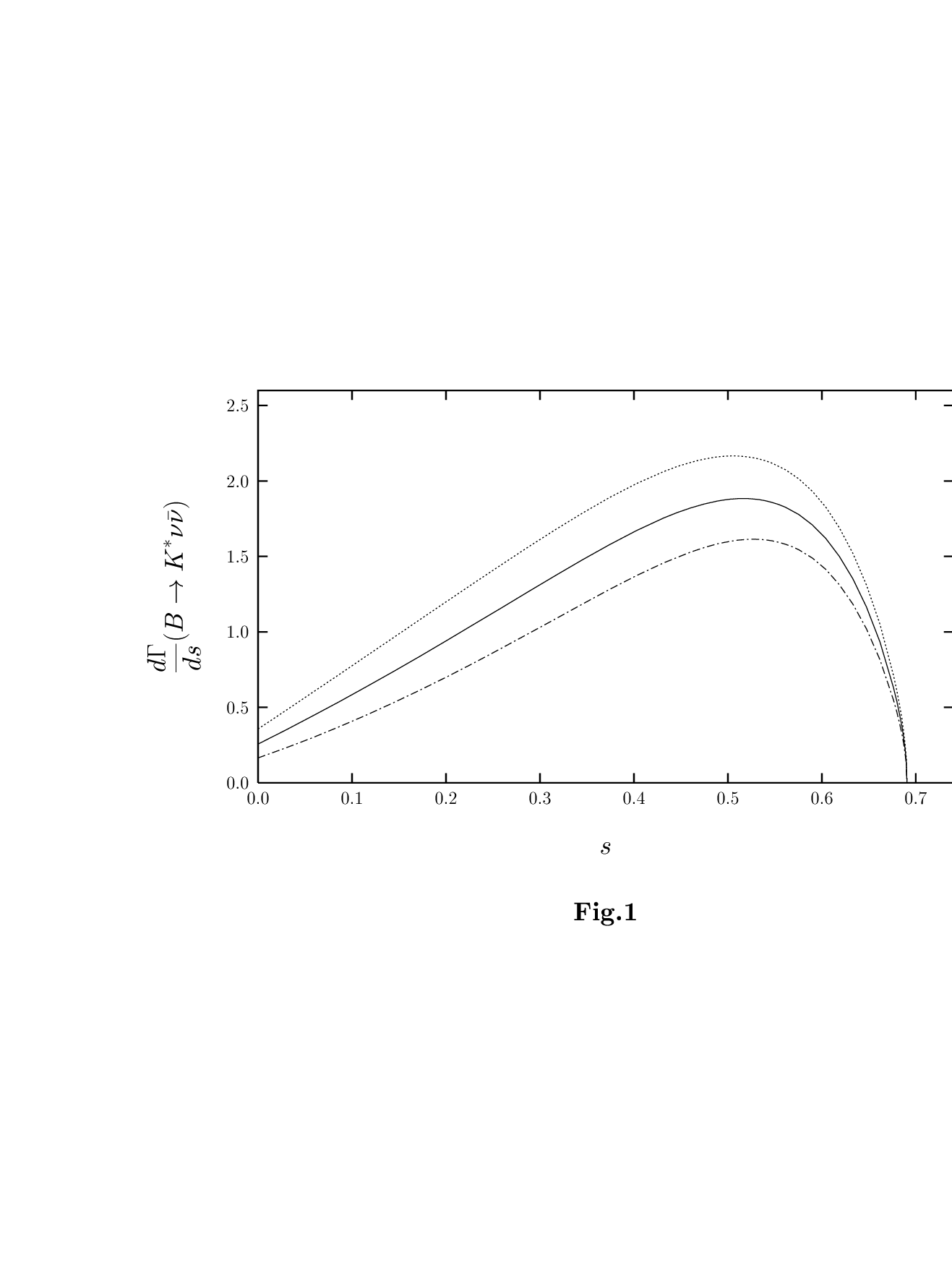,height=18cm,width=17cm,angle=0}}
\vspace*{-5cm}
\caption{\em Differential decay width  
${d \Gamma \over d s}(B \rightarrow K^* \nu \bar \nu)$ 
as a function of  the normalized momentum transfer square, 
$s \equiv q^2/m_{_B}^2$, in units of 
$\left[ 8.84 \times 10^{-18}~ \left(\displaystyle{\frac{\left | V_{tb}
V_{ts}^*\right |}{0.045}} \right)^2 \right]$ {\rm GeV}.
Dotted and dash-dotted curves correspond to the cases when
the uncertainty is added and subtracted from the central values of all
form-factors, respectively. }
\end{figure}

Now we consider the differential decay widths, $
\frac{d{\Gamma}}{dq^2}(B\rightarrow \rho,K^* + \nu + \bar\nu )$.  For
the hadronic form-factors we have used the results of the works
\cite{ref21}-\cite{ref23}, $i.e.$ the monopole type form-factors 
based on light cone QCD sum rules. 
The values of the form-factors at $q^2=0$ are (see also Ref. \cite{ref26}):
\begin{eqnarray} 
A_1^{B \rightarrow K^*}(0) &=& 0.36\pm 0.05 , \nonumber \\
A_2^{B \rightarrow K^*}(0) &=& 0.40\pm 0.05 , \nonumber \\
V^{B \rightarrow K^*}(0) &=& 0.55\pm 0.08 , \nonumber \\
A_1^{B \rightarrow {\rho}}(0) &=& 0.30\pm 0.05 , \nonumber \\
A_2^{B \rightarrow {\rho}}(0) &=& 0.325\pm 0.05 , \nonumber \\
V^{B \rightarrow {\rho}}(0) &=& 0.37\pm 0.07 , \nonumber \\
f_+^{B \rightarrow K }(0)&=& 0.29\pm 0.05 , \nonumber \\
{\rm and}~~~f_+^{B \rightarrow \pi }(0)&=& 0.32 \pm 0.05 . 
\label{(16)}
\end{eqnarray}
Note that all errors, 
which come from the uncertainties of the $b$ quark mass,
the Borel parameter variation, wave functions, non-inclusion of higher
twists and radiative corrections, are added in quadrature.

%
\begin{figure}[tb]
\vspace*{-5cm}
\hspace*{-2cm}
\centerline{\epsfig{figure=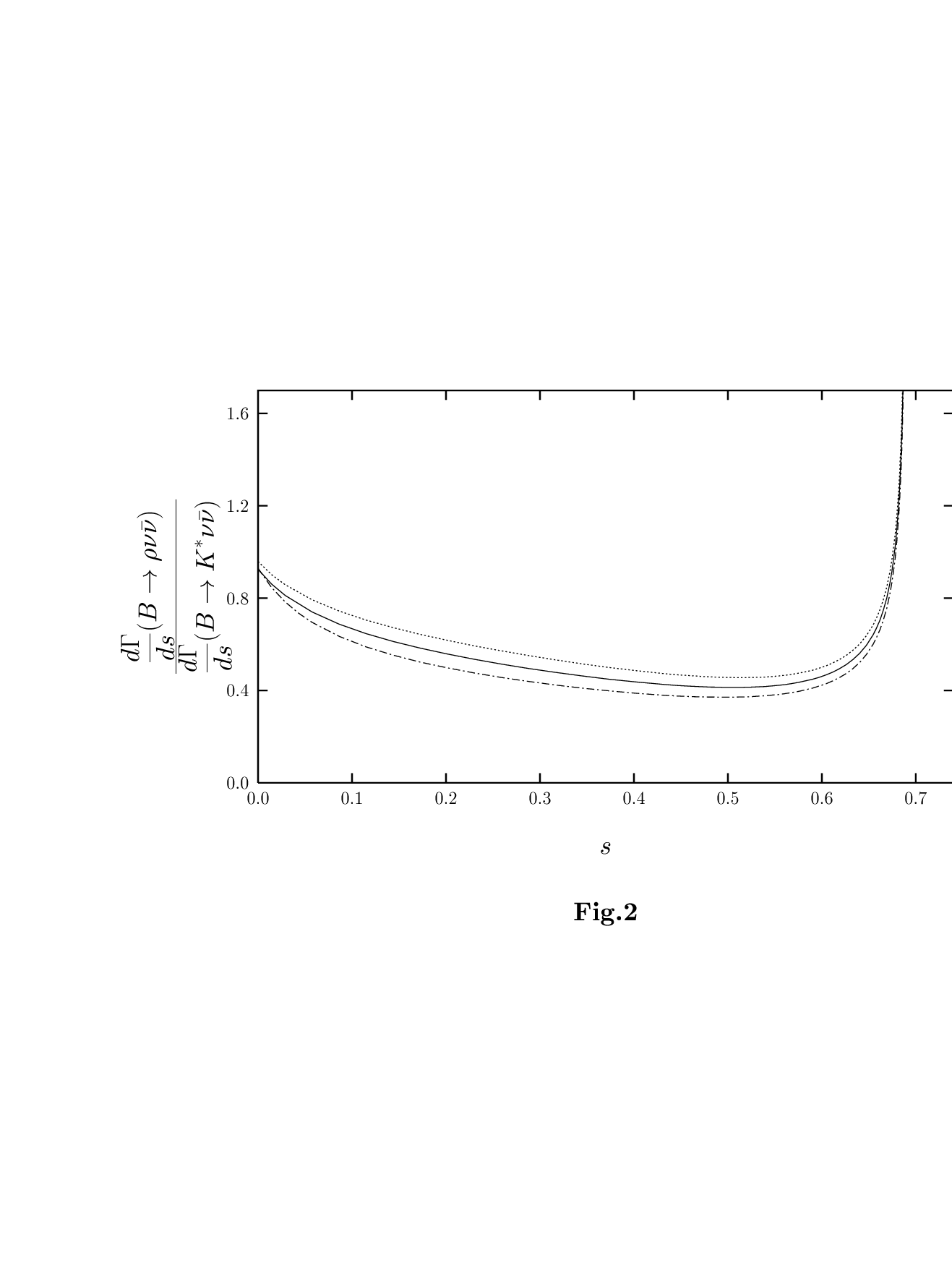,height=18cm,width=17cm,angle=0}}
\vspace*{-5cm}
\caption{\em Ratio of the differential decay rates 
$B \rightarrow \rho \nu \bar\nu $ and $B\rightarrow K^* \nu \bar\nu$, 
in units of $\left | \frac{V_{t d }}{V_{ts}} \right |^2 $, 
as a function of the normalized momentum transfer square, 
$s \equiv q^2/m_{_B}^2$.}
\end{figure}

In Fig. 1, we present the differential decay width $ d{\Gamma} /
{ds}(B\rightarrow K^*\nu \bar\nu )$ as a function of the normalized
momentum transfer square, $s \equiv q^2/m_{_B}^2$.  In Fig. 2, we show
the $q^2$ dependence of the ratio of the differential decay rates
$B\rightarrow \rho \nu \bar\nu $ and $B\rightarrow K^* \nu\bar\nu$,
normalized to $\left | \frac{V_{t d }}{V_{ts}} \right |^2 $.
In these figures, dotted and dash-dotted curves correspond to the
cases when the uncertainty is added and subtracted from the central values
of all form-factors, respectively. For the central solid curve we
use the central values of form-factors. We note that the errors
in the differential decay width of Fig. 1 due to the form-factors
uncertainties are about $\sim \pm 20\%$.  However, the errors in the
ratio of Fig. 2 are reduced to about $\sim \pm 10\%$.  We
conclude that even though the errors from uncertainties of the
form-factors for each channel are substantial, those in the
corresponding ratio are comparatively small, and that for precise
determination of the elements of the CKM matrix the investigation of
the corresponding ratio is very suitable.
We also note that the uncertainties for our main results, Eqs. (12)--(16),
where we only assume flavor $SU(2)$ (isospin),
should be even much smaller than that shown in Fig. 2, since
there we had to assume flavor $SU(3)$ symmetry.

For completeness we present the integrated value for the branching fractions 
of $B\rightarrow K^* \nu \bar\nu$ and $B\rightarrow K \nu \bar\nu$
as well as the value of the ratio
${\cal B}(B\rightarrow \rho \nu \bar\nu ) / 
{\cal B}(B\rightarrow K^* \nu \bar\nu)$ 
and
${\cal B}(B\rightarrow \pi \nu \bar\nu ) / 
{\cal B}(B\rightarrow K \nu \bar\nu)$:
\begin{eqnarray}
{\cal B}(B\rightarrow K^* \nu \bar\nu) &=& 1.7\times (1\pm 0.16)\cdot 10^{-5} 
\left |
\frac{V_{ts}V_{tb}}{0.045} \right | ^2 , \nonumber\\
\frac{{\cal B}(B\rightarrow \rho \nu \bar\nu )}
{{\cal B}(B\rightarrow K^* \nu \bar\nu)} &=&
 0.52 \times (1\pm 0.1 ) \left | \frac{V_{td}}{V_{ts}}\right |^2 , \nonumber\\
{\cal B}(B\rightarrow K \nu \bar\nu) &=& 7.8\times (1\pm 0.25)\cdot 10^{-6} 
\left | \frac{V_{ts}V_{tb}}{0.045} \right | ^2 , \nonumber\\
{\rm and}~~~\frac{{\cal B}(B\rightarrow \pi \nu \bar\nu )}
{{\cal B}(B\rightarrow K \nu \bar\nu)} &=&
 1.29 \times (1\pm 0.2 ) \left | \frac{V_{td}}{V_{ts}}\right |^2 .
\end{eqnarray}
The values of the main input parameters, which appear in the expressions for
the decay widths are
\begin{displaymath} 
m_b=(4.8 \pm 0.1)~{\rm GeV}, \quad m_{\rho} \approx 0.77~ {\rm GeV}, \quad
m_{_{K^*}}=0.892~ {\rm GeV}. 
\end{displaymath}
For the $B$ meson life time, we take 
$\tau (B_d)=1.56 \cdot 10^{-12}~ sec$ \cite{ref25}. 

\section{Discussions and conclusions} 
 
We proposed a new method for the precise determination of
$ \left | \frac{V_{td}}{V_{ub}} \right | $ from the 
ratios of the branching fractions
\begin{displaymath} 
{\cal R}_\rho=\frac{{\cal B}(B \rightarrow \rho \nu \bar\nu )}
{{\cal B}(B\rightarrow \rho \nu e)}~~~{\rm and}~~~
{\cal R}_\pi=\frac{{\cal B}(B \rightarrow \pi \nu \bar\nu )}
{{\cal B}(B\rightarrow \pi e \nu )} .
\end{displaymath}
As is well known, each partial decay width depends very strongly
on hadronic form-factors.
However, as also shown in Eqs. (9)--(13), these ratios, 
${\cal R}_\rho, {\cal R}_\pi$, are free of
any hadronic uncertainties, if small isospin breaking effects are
neglected.  Measurements of ${\cal R}_{\rho,\pi}$ allow to determine 
$ \left | \frac{V_{td}}{V_{ub}} \right | $ with little theoretical error, 
which equals $ ( \frac{\sin \gamma }{\sin \beta } ) $ 
for the CKM version of CP-violation within the Standard Model. 
Therefore, ${\cal R}_{\rho,\pi}$
measures a relation between two different phases angles, which can be
measured separately by experiments. We also found that each
exclusive channel $B\rightarrow (K, K^*, \rho , \pi ) \nu \bar\nu $
has rather large theoretical uncertainties due to the unknown hadronic
form-factors, as shown in Fig. 1. In order to reduce these
uncertainties we have considered the ratio of the corresponding
exclusive channels, $e.g.$ $(B\rightarrow \rho \nu
\bar\nu)/(B\rightarrow K^* \nu \bar\nu)$, as shown in Fig. 2.

A few words about experimental statistics for detecting the 
$ B \rightarrow \rho \nu \bar\nu $ decay follow: 
Future symmetric and asymmetric 
$B$ factories should produce much
more than $ \sim 10^9$ $B-\overline{B}$ mesons by the year 2010.  
With $10^9$~ $B$ mesons effectively reconstructed, 
the number of expected events 
for $ B^{\pm}\rightarrow {\rho}^{\pm}\nu \bar\nu $ channel is
\begin{displaymath} 
N \equiv {\cal B}( B \rightarrow {\rho} \nu \bar \nu) 
\times 10^9 \sim 100 ~~~({\rm and}~~
N( B \rightarrow K^* \nu \bar\nu) \sim 2 \times 10^4).
\end{displaymath} 
And the statistically estimated error for 
$B \rightarrow {\rho} \nu \bar \nu$ decay is approximately
\begin{displaymath} 
\frac{1}{\sqrt{N}} \approx \frac{1}{10} = 10 \% . 
\end{displaymath}
We argue that within the next decade the decay channel 
$ B^{\pm}\rightarrow {\rho}^{\pm}\nu \bar\nu $
has a good chance for being detected in future $B$ factories.

Note that the inclusive channels $B \rightarrow X_{d,s} \nu \bar\nu $
are also free of any theoretical uncertainties.  However, measuring
inclusive channels in experiments would be very difficult because of
the two missing neutrinos and (many) hadrons.  
For completeness, we give here the
summarized results for the inclusive decays in the lowest order:
\begin{eqnarray} 
\frac{{\cal B}(B \rightarrow X {\nu} \bar\nu )}
{{\cal B}(B \rightarrow X e^- \nu )}
\approx \frac{{\cal B}(B \rightarrow X_s {\nu} \bar\nu )}
{{\cal B}(B \rightarrow X_c e^- \nu )}
={3} \left | \frac{V_{ts}}{V_{cb}} \right |^2  C ,
\end{eqnarray}
\begin{displaymath} 
 \frac{{\cal B}(B \rightarrow X_d {\nu} \bar\nu )}
{{\cal B}(B \rightarrow X_u e^- \nu )}
={3} \left ( \frac{\sin {\gamma}}{\sin {\beta}} \right )^2  C ,
\end{displaymath}
and
\begin{eqnarray} 
{\cal B}(B \rightarrow X_s {\nu} \bar\nu)
&\sim& 3 \times 10^{-5}, \nonumber \\
{\cal B}(B \rightarrow X_d {\nu} \bar\nu)
&\sim& 5  \times 10^{-7}. \nonumber
\end{eqnarray}
In derivation of Eq. (20) we have neglected the charm quark mass.
\\

\noindent
{\Large \bf Acknowledgements}\\
One of the authors (T.M.A.) sincerely thanks Mustafa Savc{\i} for
helpful discussions and for his assistance in numerical calculations.
We thank M. Drees for careful reading of the manuscript and his
valuable comments.  The work of CSK was supported in part by the CTP
of SNU, in part by the BSRI Program BSRI-97-2425, in part by
Non-Directed-Research-Fund of 1997, in part by Yonsei University
Faculty Research Fund of 1997, and in part by the KOSEF-DFG, Project
No. 96-0702-01-01-2.

\end{document}